\documentclass[useAMS,usenatbib,onecolumn,usegraphicx]{mn2e}
\title[Evolution of initially localized perturbations in stratified ionized disks]{Evolution
of initially localized perturbations in stratified ionized disks}
\author[Edward Liverts and Michael Mond]{Edward Liverts\thanks{E-mail:
eliverts@bgu.ac.il (EL); mond@bgu.ac.il (MM)} and Michael Mond \\
Department of Mechanical Engineering,  Ben-Gurion
University of the Negev, \\ P.O. Box 653, Beer-Sheva 84105,
Israel}
\begin{document}

\date{Accepted ---. Received ----; in original form ----}

\pagerange{\pageref{firstpage}--\pageref{lastpage}} \pubyear{}

\maketitle

\label{firstpage}

\begin{abstract}
A detailed solution of an initial value problem of a vertically localized
initial perturbation in rotating magnetized vertically stratified disk is
presented. The appropriate linearized MHD equations are solved by
employing the WKB approximation and the results are verified numerically.
The eigenfrequencies as well as eigenfunctions are explicitly
obtained. It is demonstrated that the initial perturbation remains confined within
the disk. It is further shown that thin enough disks are stable but as their
thickness grows increasing number of unstable modes participate in the
solution of the initial value problem. However it is demonstrated that due to
the localization of the initial perturbation the growth time of the
instability is significantly longer than the calculated inverse growth rate
of the individual unstable eigenfunctions.
\end{abstract}


\section{Introduction}

Magnetohydrodynamic (MHD) instabilities play a major role in a great
variety of astrophysical and space applications. Their importance is
epitomized by the magneto-rotational instability (MRI) that has been
discovered by \cite{velikhov} and \cite{chandra} for infinite cylinders and
rediscovered by \cite{bh} in astrophysical context.
It is widely believed to be one of the prime candidates to provide a viable clue
to solving the age old puzzle of the outwards transfer of angular momentum in a
plethora of astrophysical disk configurations.

However, for a case of finite size system, the common practice in stability analysis
to expand the state vector of the linearized dynamical system
(the MHD system of equations in our case) in plane waves as:
${\bf u}(z,t)={\bf A}e^{-i\omega t+ikz}$,
where $z$ is the spatial coordinate (for simplicity only one spatial
coordinate will be considered), $k$ is the wave number, and $\omega$
is the natural frequency of the system, is not applicable especially for
inhomogeneous systems. Instead, for such cases the appropriate boundary value
problem (BVP) for a given set of initial conditions (IC) should be solved
(see for example the well-known work of \cite{landaupol}). The solution of such
problems is facilitated by obtaining the natural frequencies of the bounded system
$\omega_n$ for which the state vector of the linearized dynamical system may be written as
${\bf u_n}(z,t)={\bf A}_n(z)e^{-i\omega_n t}$, where ${\bf A}_n(z)$
are the eigenfunctions of the BVP subjected to specific boundary conditions (BC)
(see also \cite{sm,cc,ck}). The main purpose of the current work is to employ such
approach in order to investigate the dynamical evolution of localized initial
perturbations in rotating magnetized disks of finite thickness.

In reality, small perturbations, especially in a system of finite
dimensions, do not have the form of a single monochromatic wave, but
are rather a superposition of individual waves, i.e., wave packets.
Furthermore, the asymptotic behavior at large times of the wave
packet may significantly differ from that of any one of its
individual components. Thus, under certain circumstance, even though
some of the components of the wave packet may individually grow with
time without bound, the wave packet as a whole can remain bounded at
a given place and even decay to zero as the packet is convected
away. In such cases the initial perturbations give rise to what is
defined as convective instabilities. In other cases, the
exponentially growing components of the wave packets may indeed
cause the perturbation to grow without bound at each place. In such
cases, the perturbations are defined as absolute instabilities.
Furthermore, the BC could significantly change the
results of the stability analysis (global stability). It is therefore of utmost importance
to investigate the asymptotic (in time) development of the solutions
of the appropriate Cauchy problems in order to determine whether the
system is stable or not.
%Ascertaining that some of the individual
%monochromatic waves grow without bound provides a necessary
%condition for instability but, as we shall see, certainly not a
%sufficient one.

Various sets of rules in order to distinguish between convective and
absolute instabilities have been given by
\cite{landau,sturrock,fks,akhiezer,lp,drazin}, while \cite{Huerre
Monkewitz} have reviewed more recent developments of the theory
pertaining to hydrodynamic stability. The importance of studying
the influence of a finite system size on stability analysis
has been demonstrated by \cite{budker,sturrock} who reported that the size of system
could play a stabilizing role for the two stream instability.

%The importance of the
%distinction between convective and absolute instability has been
%demonstrated in the Riga dynamo experiment in which convective
%instabilities were converted into absolute instabilities by an
%appropriate redesign of the basic flow \cite{Gailitis}.
One should note also that such
understanding may play an important role in recent attempts to
observe the MRI in the laboratory \citep{Colgate,Rudiger,Goodman}.

\section{The magneto rotational instability}

\subsection{Formulation of the problem and formal solution}
The effect of the finite size of the system is investigated by re-examining
the MRI for the case of rotating disks with finite thickness.
The basic MHD equations that describe the dynamical development of
the system are:
\begin{equation}
\rho\frac{d\vec{V}}{dt}=-\vec{\nabla}P+\frac{1}{c}\vec{J}\times\vec{B}+\rho\vec{G},
\quad \vec{\nabla}\cdot\vec{V}=0,
\label{ind}
\end{equation}
\begin{equation}
\frac{\partial\vec{B}}{\partial t}=-c\vec{\nabla}\times\vec{E},
\quad \vec{\nabla}\times\vec{B}=\frac{4\pi}{c}\vec{J},
\quad \vec{\nabla}\cdot\vec{B}=0,
\label{mom}
\end{equation}
where $\vec{G}$ is acceleration due to gravity, and $c$ is speed of light and
the rest of the variables have their usual meanings.
The expression for electric field $\vec{E}$ in ideal plasmas is given by:
\begin{equation}
\vec{E}=-\frac{1}{c}\vec{V}\times\vec{B}.
\label{ohm}
\end{equation}
In order to simplify the
calculations (and with no loss of generality of the results) we
assume that the Brunt-V\"{a}is\"{a}l\"{a} frequency is small in
comparison to all other characteristic frequencies in the system and
can hence be neglected. Thus, the linearized MHD system of equations
that describe radially independent perturbations in a Keplerian disk
under the influence of a constant axial magnetic field, may be represented
in the following way:

\begin{equation}
\frac{\partial {\bf u}}{\partial t}-P\frac{\partial {\bf
u}}{\partial z}+Q{\bf u}=0, \label{linearmhd}
\end{equation}
where ${\bf u}(z,t)=  ( v_r, v_{\varphi}, b_r, b_{\varphi} )^T$,
\begin{equation}
P=\left [
\begin{array}{cccc}0& 0 &\frac{\rho_0}{\beta\rho(z)} &0
 \\0 &0 &0& \frac{\rho_0}{\beta\rho(z)} \\ 1 &0& 0& 0 \\ 0& 1& 0& 0 \end{array} \right ],\;\;\;
 Q=\left [ \begin{array}{cccc}0& -2 &0 &0
 \\\frac{1}{2} &0 &0& 0 \\ 0 &0& 0& 0 \\ 0& 0& \frac{3}{2}& 0 \end{array} \right
 ],
\end{equation}%

where $\rho(z)$ is the steady state density profile, and $v_r, v_{\varphi}, b_r$, and
$b_{\varphi}$ are the perturbed
radial and azimuthal velocities ($ v_r, v_{\varphi}\ll \Omega r)$, and the radial and azimuthal
components of the magnetic field ($ b_r, b_{\varphi}\ll B_z)$, respectively. In eq.
(\ref{linearmhd}) the density is normalized to its value at the disk midplane $\rho_0=\rho(z=0)$,
the velocities are normalized to $\sqrt{2}c_s$ ($c_s$ is the sound velocity), the time is normalized
to the inverse steady state angular velocity $\Omega$, the magnetic field is scaled with the steady state
axial magnetic field $B_z$. Finally, lengths are scaled by $\sqrt{\beta}V_A/\Omega$ where
$V_A=B_z/\sqrt{4\pi\rho_0}$ is the Alfv\'{e}n velocity and $\beta$ is the familiar plasma
parameter given by $\beta=2c^2_s/V^2_A$.

As a specific example consider a thin isothermal Keplerian disk.
In order to model the finite thickness of the disk the following profile
$$\rho(r,z)=\rho_0(r)\exp(-z^2)$$
is assumed for the density. Due to the independence of the perturbations on the radial direction
$r$ is from now on merely a parameter. This is tantamount to the local approximation in $r$ with $k_r=0$,
which according to \cite{bh} is the most unstable case. For that case, assuming
that the IC are ${\bf u}(z,0) =  ( v_r(z,0), 0, 0, 0)$,
and employing the Laplace transform, the set of eqs.(\ref{linearmhd})
is reduced to the following single inhomogeneous ordinary differential equation (ODE):
\begin{equation}
L[\widetilde{b}_{\varphi}]=2i\omega\beta^2 v'_r(z,0)
\label{bphieinh}
\end{equation}
where

\begin{equation}
L[\widetilde{b}_{\varphi}]=\frac{d}{dz}\left[e^{z^2}\frac{d^2}{dz^2}\left(e^{z^2}\frac{d\widetilde{b}_{\varphi}}{dz}\right)\right]+
(3\beta+2\lambda^2)\frac{d}{dz}\left(e^{z^2}\frac{d\widetilde{b}_{\varphi}}{dz}\right)
+\lambda^2(\lambda^2-\beta)\widetilde{b}_{\varphi},
%\label{bphieinh}
\end{equation}

$\lambda=\sqrt{\beta}\omega$ represents the re-normalized spectral parameter $\omega$, and
$\widetilde{b}_{\varphi}$ is the Laplace transform of the azimuthal component of the perturbed
magnetic field. The solution of eq.(\ref{bphieinh}) should satisfy the following BC:

\begin{equation}
\widetilde{b}_{\varphi}(\pm\infty)=0, \quad
(e^{z^2}\widetilde{b}'_{\varphi})'_{|_{z=\pm\infty}}=-2i\sqrt{\beta}\lambda\widetilde{b}_r(\pm\infty)-
\lambda^2\widetilde{b}_{\varphi}(\pm\infty) = 0.
\label{mhdbcinh}
\end{equation}

Boundary conditions (\ref{mhdbcinh}) are obtained due to the requirement that at $z\rightarrow\pm\infty$,
where $\rho\rightarrow 0$ and $B_z=const$, the energy flux of the perturbation is finite.
%One should note here again that due to the density inhomogeneity the eigenfunctions of this
%problem are not plane waves $exp[i(\omega t-kz)]$.

The solutions of the BVP (\ref{bphieinh})-(\ref{mhdbcinh}) can be constructed
by the countable set of the solutions of the homogeneous equation $L[\widetilde{b}_{\varphi}]=0$,
subject to homogeneous boundary conditions. The solutions of that problem are termed eigenfunctions
of the homogeneous equation.
Assuming completeness and orthogonal property of the eigenfunctions allow to express the solution
of eq.(\ref{bphieinh}) in a simple way for an arbitrary IC by using Green's function (see appendixes
A and B for proof of orthogonality, and construction of the Green's function). The latter
is represented by an appropriate set of eigenfunction $\widetilde{b}_{\varphi, n}(z)$. Thus,
carrying out the inverse Laplace transform, the solution of eq.(\ref{linearmhd}) for $t\geq 0$
is given by the following expression:

\begin{equation}
b_{\varphi}(z,t)=4\sum_nd_n\frac{\widetilde{b}_{\varphi, n}(z)}{\omega^2_{-}(n)-\omega^2_{+}(n)}
\sin(\frac{\omega_{+}(n)-\omega_{-}(n)}{2}t)\sin(\frac{\omega_{+}(n)+\omega_{-}(n)}{2}t)
\label{disksol}
\end{equation}

where $\widetilde{b}_{\varphi, n}(z)$ are the eigenfunctions of the homogeneous equation,
$d_n$ are the coefficients of the expansion of $v'_r(z,0)$
in the complete set $\widetilde{b}_{\varphi, n}(z)$, and $\omega_\pm(n)$
are the natural frequencies that correspond to the eigenfunctions $\widetilde{b}_{\varphi, n}(z)$.

%We start to solve the BVP (\ref{bphieinh})-(\ref{mhdbcinh})
%by looking for asymptotic solutions of eq.(\ref{bphieinh})
%for $\lambda\gg1$. First we find the
%eigenfunctions of the homogenous part of the ODE given by eq.(\ref{bphieinh}).
%Under the assumption $\lambda\gg1$
\subsection{WKB solution}

We turn now to obtaining asymptotic solutions for $\widetilde{b}_{\varphi, n}(z)$ in the
limit $\lambda\gg1$ [see \citep{em} as an example of employing the WKB approximation for some problems
in inhomogeneous plasmas]. In that case the eigenfunctions may be represented by
the following version of the WKB approximation:

\begin{equation}
\widetilde{b}_{\varphi,k}=e^{-z^2}\phi_k(z)=e^{-z^2}e^{S_k(z)},\quad k=1,\ldots,4.
\label{asympt}
\end{equation}

Inserting expression (\ref{asympt}) into the homogenous part of eq.(\ref{bphieinh}) yields:

\begin{equation}
S'^4+\lambda^2\left\{3P(z,\lambda)S'^2+2Q(z,\lambda)S'+R(z,\lambda)+3S''\left[P(z,\lambda)+
\frac{2S'^2+S''}{\lambda^2}\right]\right\}+4S'S'''+S^{IV}=0
\label{eikonal}
\end{equation}

where

$$P(z,\lambda)=\frac{1}{3}\left[e^{-z^2}(2+3\beta/\lambda^2)-(6+4z^2)/\lambda^2\right],\quad Q(z,\lambda)=
-z\left[e^{-z^2}(2+3\beta/\lambda^2)+4/\lambda^2\right]$$
and
$$R(z,\lambda)=e^{-2z^2}(\lambda^2-\beta)-2e^{-z^2}(2+3\beta/\lambda^2).$$
By keeping the leading terms in $\lambda$, eq.(\ref{eikonal}) can be written as

\begin{equation}
S'^4+\lambda^2\left[3P(z,\lambda)S'^2+2Q(z,\lambda)S'+R(z,\lambda)\right]=0
\label{eikonal0}
\end{equation}

The solutions of eq.(\ref{eikonal0}) provide valid approximations for the eigenfunctions
of $L$ throughout the disk except in the vicinity of the turning points.
Assuming that the latter occur at $|z_0|\gg 1$, it can be shown that the last term
in eq.(\ref{eikonal0}) becomes exponentially small in comparison with other terms
(this assumption will be validated later on). In that case the equation that determines the turning points
is given by
\begin{equation}
\frac{\textsf{D}(z_0,\lambda)}{\lambda^4}=\lambda^2P^3(z_0,\lambda)+Q^2(z_0,\lambda)=0.
\label{discr}
\end{equation}
Asymptotic solution of eq.(\ref{discr}) for large $\lambda$ yields the following expression for the turning points:
\begin{equation}
z_0=\pm\sqrt{\ln \frac{2\lambda^2}{\ln 2\lambda^2}} + O\left(z_0^{-3}\right),
\end{equation}
as well as the following form of the discriminant in the vicinity of the turning points:
\begin{equation}
\textsf{D}(z,\lambda)\approx K(z-z_0), \quad K=\frac{d D(z,\lambda)}{dz}|_{z=z_0}
\approx-6z^7_0(1+5z^{-2}_0)
\label{expan}
\end{equation}
In the outer regions namely $z < -z_0$ and $z > z_0$
all four functions $S_k(z)$ are real, whereas in the inner region $-z_0 < z < z_0$ the solutions for $S_k(z)$
acquire non zero imaginary part. Thus, the turning points separate
the inner range $-z_0 < z < z_0$ where the eigenfunctions are oscillatory from the outer regions $z < -z_0$
and $z > z_0$ where the eigenfunctions decrease exponentially away from $z_0$ and
have no zeros at a finite distance $z$. The behavior of the various solutions of eq.(\ref{eikonal0})
as well as the turning points may be seen in Fig.1.

\begin{figure*}
  %\vspace*{174pt}
\includegraphics[width=120mm]{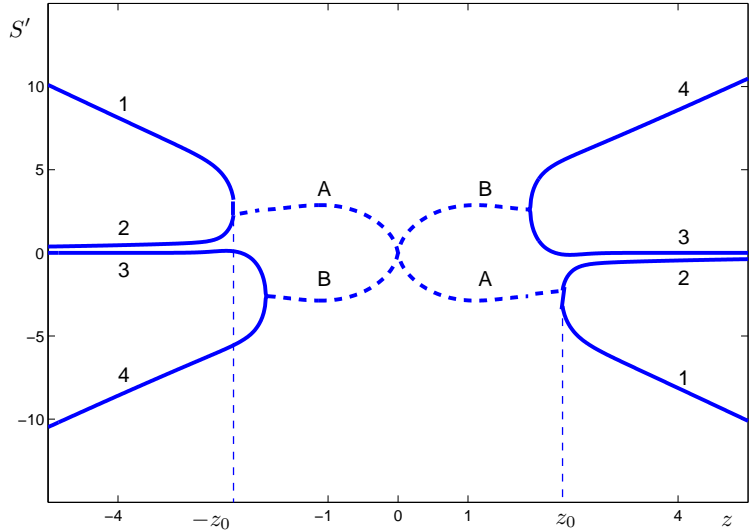}
  \caption{The solutions of eq.(\ref{eikonal0}), obtained numerically
for $\beta=10,\lambda=25$. Each of the dashed lines ($A,B$) represents the same real part of two complex
conjugate solutions.}
\label{fig1}
\end{figure*}

It is obvious from assumption (\ref{asympt}) and BC (\ref{mhdbcinh}) that only solution 1 is admissible.
The turning points are obtained as the merging of solution 1 and 2. In the region between the turning points
two solutions are admissible, that are complex conjugates of each other. Again, due to BC (\ref{mhdbcinh})
those solutions are marked by A that depicts their common real part.

In the vicinity of the turning points
the solutions assumed in eq.(\ref{asympt}) are invalid and instead an appropriately re-scaled eq.(\ref{bphieinh})
is solved. That solution is then matched asymptotically to the outer solution given
by (\ref{asympt}) and (\ref{eikonal0}) where $S'$ is obtained from branch 1 in Fig.1.

In order to obtain
an asymptotic expression for $S'$ in branch 1 it is noticed again that $R(z,\lambda)$ is asymptotically
much smaller than the rest of the terms that multiply $\lambda$ in eq.(\ref{eikonal0}). As a result,
one of the roots is asymptotically zero (branch 3) while the rest three roots are obtained from
the Cardano solution of the reduced eq.(\ref{eikonal0}). Thus, the root that is represented by branch 1 (the only
admissible root) gives rise to the following solution in the outer region close to the turning point:
\begin{equation}
\phi(z)=\frac{C}{2\sqrt[4]{|K|}}\frac{1}{(z-z_0)^{\frac{1}{4}}}\exp{[-\int_{z_0}^z\sqrt{K(z_0-\xi)}d\xi]}
\label{right}
\end{equation}
In the inner region the solution near the turning points is given by
\begin{equation}
\frac{1}{\sqrt[4]{|K|}(z_0-z)^{\frac{1}{4}}}\left\{C_1\exp{[i\int_{z_0}^z\sqrt{K(\xi-z_0)}d\xi]}+
C_2\exp{[-i\int_{z_0}^z\sqrt{K(\xi-z_0)}d\xi]}\right\}.
\label{left}
\end{equation}
The coefficients ($C_{1,2}$) are determined by the connection formulas that express the asymptotic matching
of the solution found to the left (\ref{left}) and to the right (\ref{right}) of a turning point $z_0$. Following
\cite{landauQM} the connecting relations are:
$$C_{1,2}=\frac{C}{2}\exp{(\pm i\frac{\pi}{4})}.$$
Applying the same rule to the region close
to the left turning point located at $-z_0$ and requireing that the two expressions are the same
throughout the region $-z_0<z<z_0$ (the sum of their phase must be multiple of $\pi$)
results in the following expression for the eigenfunction:
\begin{equation}
\widetilde{b}_{\varphi, n}(z,\omega_\pm(\beta,n))=\cos\left[\kappa\Phi(z)-\frac{\pi}{4}\right]
e^{-\frac{z^2}{4}},
\label{wkbsol}
\end{equation}
where $\Phi(z)=\int^{z_0}_z e^{-\frac{\zeta^2}{2}}d\zeta$. The analytical expression (\ref{wkbsol}) for the solution
in the inner region has been obtained by noticing that away from the turning points $S'Q(z,\lambda)$ is smaller
than the rest of the terms in eq.(\ref{eikonal0}) and hence to leading order the latter is a biquadratic equation.
The  discrete set of eigenvalues $\omega$ is now determined from the Bohr-Sommerfeld relation that is given by:
\begin{equation}
\kappa=\sqrt{\frac{\beta}{2}}\sqrt{3+2\omega_\pm(\beta,n)^2+\sqrt{9+16\omega_\pm(\beta,n)^2}}=
\frac{\pi}{\Phi(-z_0)}\left(n+\frac{1}{2}\right),
\label{BZ}
\end{equation}
The integer $n$ in the last expression describes the number of zeros of the solution
within the inner region [see eq.(\ref{wkbsol})], whereas $\omega_\pm(\beta,n)$ are the natural frequencies of the system
that correspond to such eigenfunctions. Strictly speaking expression (\ref{BZ}) is valid for $n >> 1$.
Nevertheless, as is often the case for WKB-type solutions, numerical calculations
indicate that expression (\ref{wkbsol}) provides a close approximation also for $n$'s as low as 1 (see Fig.2).
It is also evident from Fig.2 that the wave length of the perturbation depends on $z$ and approaches the disk thickness
close to the turning points.

\begin{figure*}
  %\vspace*{174pt}
\includegraphics[width=74mm]{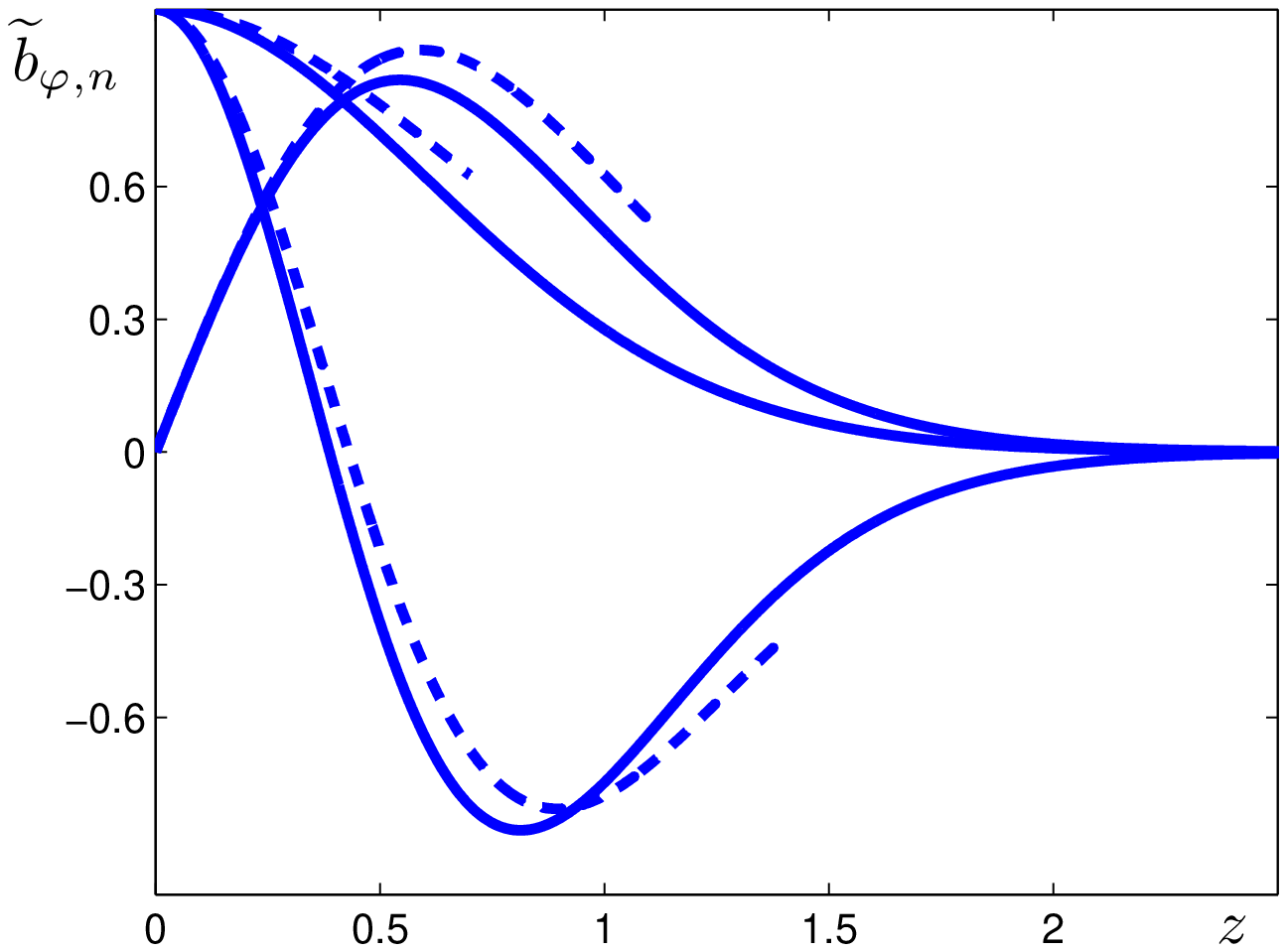}(a)
\includegraphics[width=74mm]{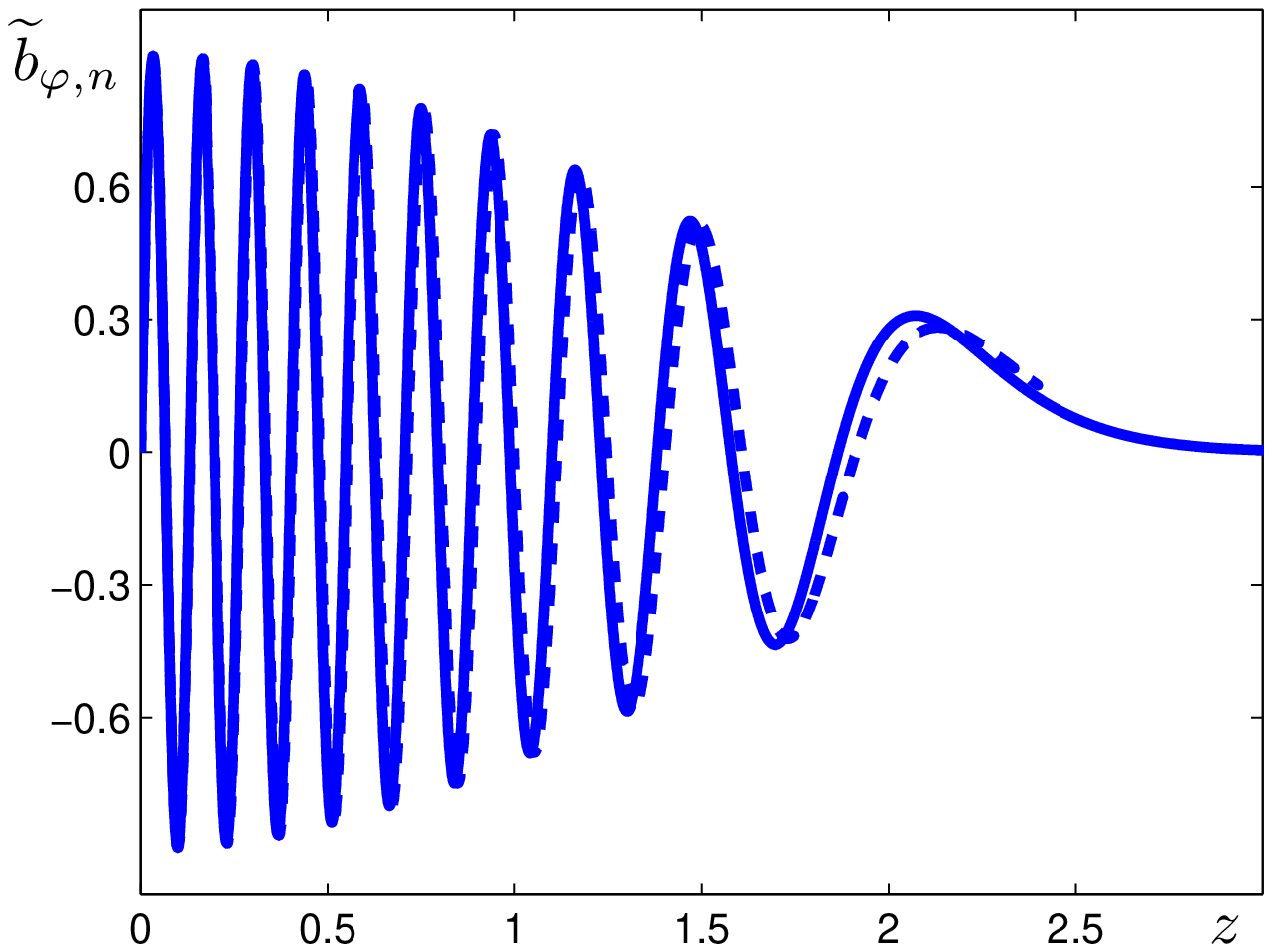}(b)
  \caption{Comparison between the WKB solutions [eq.(\ref{wkbsol}), dashed line] and the numerical solution (full line)
of the homogeneous part of eq.(\ref{bphieinh}); (a) for the first three eigenfunctions ($n=0,1,2$) and
(b) - $n=37$.}
\label{fig2}
\end{figure*}

As $n=0$ is the first excited unstable mode, it is easy to show that the disk is stable for $\beta\leq\beta_*$
where $\beta_*=\pi/(24[Erf(\frac{z_0}{\sqrt{2}})]^2)$ ($n=1,\beta_*=3\pi/(8[Erf(\frac{z_0}{\sqrt{2}})]^2)$).

\subsection{Numerical solution}

\begin{figure*}
  %\vspace*{174pt}
\includegraphics[width=120mm]{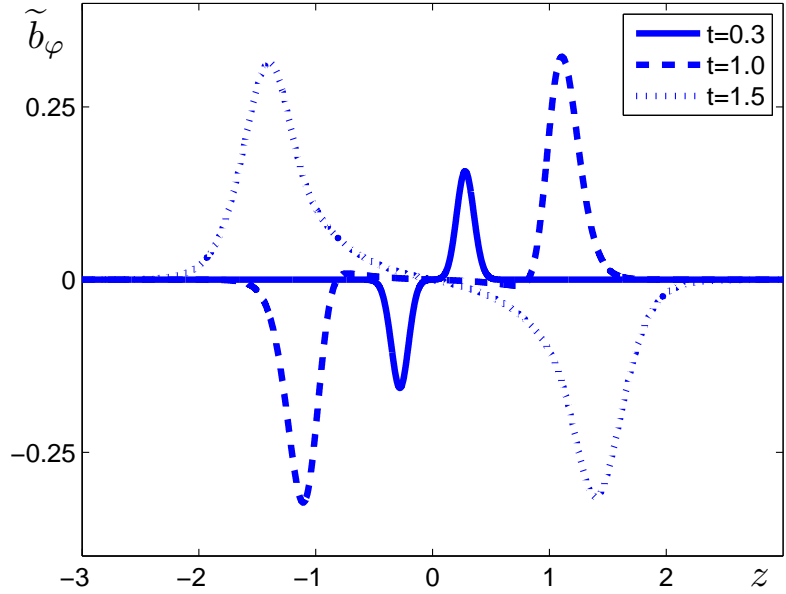}
  \caption{Time evolution of the profile of the perturbed azimuthal component of the
magnetic field. The different profiles were calculated at times
[0.3, 1.0, 1.5] for $\beta=3\pi/8$, $\Delta=0.1$ .}
\label{fig3}
\end{figure*}

To further follow the development of a localized initial perturbation governed by eq.~(\ref{linearmhd}),
consider for example an IC given as:
\begin{equation}
v_r(z,0)=e^{-z^2/\Delta ^2},
\label{gauss}
\end{equation}
and zero for the rest of the initial values of the perturbed physical variables.
It is easy to see that the perturbation is localized within the disk,
if $\Delta << 1$. The dynamical development of the localized initial perturbation that
is described by eq.(\ref{disksol}) depends on the disk "thickness" defined by the plasma
parameter $\beta$ (notice that now velocities are scaled by the Alfv\'{e}n velocity which means that the disk thickness is $2\sqrt{\beta}$ ).
If the disk is thin enough such that $\beta \leq \beta_*$ all the natural frequencies $\omega_\pm(n)$
that are obtained from eq.~(\ref{bphieinh}) are real and hence the disk is stable. In that case following
(\ref{disksol}) the profile of the perturbed azimuthal magnetic field represented by
two identical wave packets that move in opposite directions namely up and down. Upon reaching the upper
and lower turning points (located at $\pm z_0$) the two packets are reflected and continue their motion through
the disk. The amplitudes of the identical wave packets change in time like $\propto \sin(t)$. It should be noted
however that this picture is valid if the initial perturbation is localized enough. As example for such behavior,
$b_{\varphi}(z,t)$, which has been obtained by using eq.(\ref{disksol})
$\Delta =0.1$, and $\beta=3\pi/8$
is depicted in Fig. \ref{fig3}. In this particular case all eigenfunctions have real
natural frequencies [$\omega_\pm(1)=1.54,3.64, \omega_\pm(3)=3.99,6.01, ...$]
and there are no growing modes at all (the wave packets indeed
do not grow in time, just move back and forth inside the disk with local Alfv\'{e}n velocity).

For "thick" disk ($\beta > \beta_*$) some of the natural frequencies have an imaginary part. As the thickness
of the disk increases so does the number of unstable modes.
Thus for long time $b_{\varphi}(z,t)\sim\widetilde{b}_{\varphi, n}(z)exp(\mid\Im\omega_{-}(n)\mid t)$
where $n$ corresponds to natural frequency with a biggest imaginary part. However,
for sufficiently localized initial perturbation the
fastest growing mode has sufficiently small initial amplitude. The perturbation then
keeps it's original shape for a long time until a growing mode emerges out of the wide initial spectrum
and is of order of the amplitude of initial wave packet. Raising the value of $\beta$ to $10$, one unstable mode,
with growth rate $\gamma=0.749$ enters the spectrum of the eigenmodes. Employing again eq.(\ref{disksol}) together with
eigenfunctions $\widetilde{b}_{\varphi, n}(z)$ results in Fig.4. It is indeed seen that
after long time the unstable mode dominates the perturbation. Thus, the latter eventhough starting as a well
confined perturbation develops into a global instability. It should be noted however, that the time it takes
the unstable mode to emerge from the wide spectrum that makes up the localized initial perturbation is about
4 times longer than the predicted growth time (inverse growth rate). Indeed, as the initial perturbation
is more localized so the growth time of the most unstable mode is increased relative to its linear predicted value.

\begin{figure*}
  %\vspace*{174pt}[!ht]
\includegraphics[width=120mm]{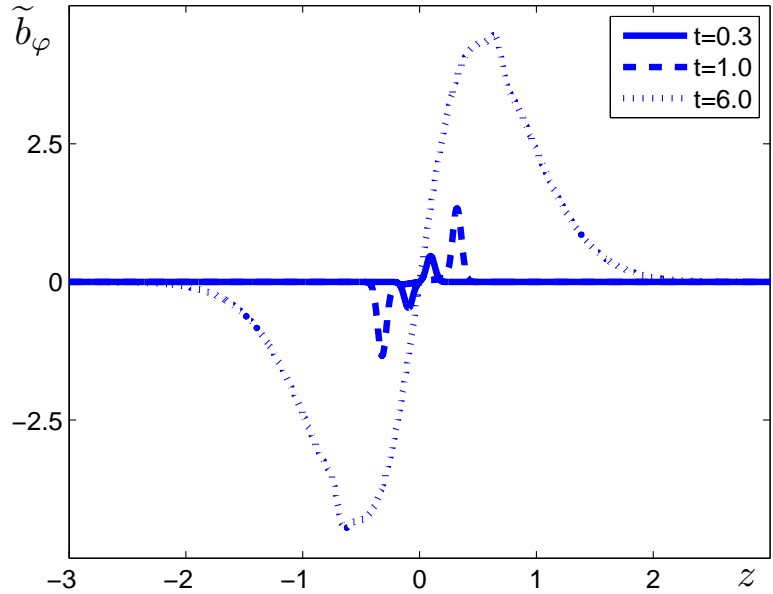}
  \caption{Long time evolution of the profile of the perturbed azimuthal component of the
magnetic field for $\beta=10$, $\Delta=0.05$. The different profiles were calculated at times
[0.3, 1.0, 6.0].}
\label{fig4}
\end{figure*}
%\begin{figure}[!ht]
%\epsscale{.80}
%\plotone{fig4_05.eps} \caption{Long time evolution of the profile of the perturbed azimuthal component of the
%magnetic field for $\beta=10$, $\Delta=0.05$. The different profiles were calculated at time
%[0.3, 1.0, 6.0]
%\label{fig4}}
%\end{figure}
It is finally instructive to plot the wave number $n^*$ of the most unstable mode as a function of $\beta$ (Fig.5).
Asymptotic estimation reveals that $n^*$ is proportional to $\sqrt{\beta}$. This implies
that the wavelength of the most unstable mode is of the order of $h/n^* \sim V_A/\Omega$ (where $h$
is the thickness of the disk) which to leading order does not depend on $\beta$.

\begin{figure*}
  %\vspace*{174pt}
\includegraphics[width=120mm]{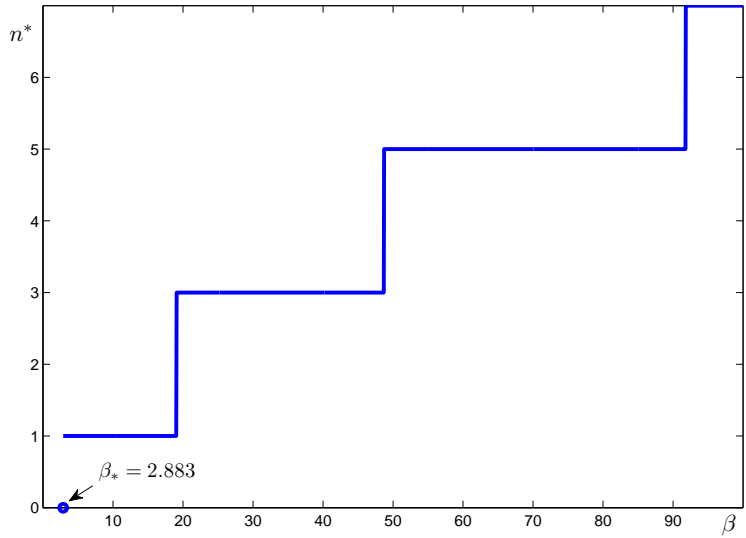}
  \caption{Axial standing wave number of the most unstable
mode as a function of the plasma beta.}
\label{fig5b}
\end{figure*}
%\begin{figure}[!ht]
%\epsscale{.80}
%\plottwo{fig5_a.eps}{fig5_b.eps}\caption{Growth rate $\gamma$ and number of zeros for the most unstable
%standing wave vs plasma beta parameter.
%\label{fig5}
%}
%\end{figure}

\subsection{Limit of infinite "thickness"}
It is instructive to examine the limit of thick disk such that effect of boundaries may be ignored.
Such an assumption is justified when the Alfv\'{e}n traverse time is longer than the time of interest
(say inverse growth rate). Renormalizing therefore lengths to $V_A/\Omega$ solution is sought for
$z\ll\sqrt{\beta}$, where $2\sqrt{\beta}$ is now the disk's thickness. It is readily seen that for such case
$\widetilde{b}_{\varphi, n}(z)\rightarrow \widetilde{b}_\varphi(k,z)=\cos kz$ for odd initial perturbations
[see eq.(\ref{wkbsol}) where $k$ is defined as $\sqrt{\pi/2\beta}(n+1/2)$] so that the eigenfunctions
can be approximated by plane waves. Thus, due to $1/\sqrt{\beta}\rightarrow0$, the sum (\ref{disksol})
is replaced by an integral over $k$ and eq.(\ref{disksol}) can be written as
\begin{equation}
b_{\varphi}(z,t)\rightarrow
\frac{1}{4\pi^2}\int_Ce^{-i \omega t}d\omega\int_{-\infty}^{\infty}\frac{{\bf A}(k)e^{i kz}dk}{D(k,\omega)}
\label{aa}
\end{equation}
where $C$ is the contour of integration in the complex $\omega$ plane that is a straight line parallel to the real
axis and passing above all singular points of the integrand, ${\bf A}(k)$ is the Fourier transform of the right hand side of
eq. (\ref{bphieinh}) and the denominator of the integrand is given by
\begin{equation}
D(\omega ,k)=(\omega ^2-k^2)^2-\omega ^2-3k^2. \label{mri}
\end{equation}
It is important to notice now that since the initial perturbation is well localized in space ${\bf A}(k)$
in contrast is close to a constant that is proportional to the localization length.
As a result it can be shown now that $b_{\varphi}(z,t)$ obtained by using (\ref{aa}) is not zero only if the
integrand has a singularities of order two (branch points) (see in details in \cite{fks,akhiezer}).
Thus, examining the equation $D(\omega ,k)=0$ [the classical MRI dispersion equation \citep{bh}]
it is clear that there is just one branch point
that gives rise to instability, which is $\omega=3i/4,k=\sqrt{15}/4$. Consequently, the long time
behavior of the perturbation is given by
\begin{equation}
b_{\varphi}(t)\sim e^{3t/4}/\sqrt{t}.
\label{longtime}
\end{equation}
As remarked above this result is due to the fact that the initial perturbation is localized in space.
If, however, (a not too physical) monochromatic perturbation is considered, ${\bf A}(k)$ is a delta
function and the familiar purely exponential growth is recovered.

\section{Conclusions}
The importance of solving the initial value problem with some appropriate boundary conditions
is highlighted.
In the classical works of \cite{velikhov,chandra}, and \cite{bh},
an infinite cylinder has been considered and the effects of the boundary conditions were neglected.
Therefore, in those works, naturally, the thickness of the disk does not play any role and
consequently cannot influence the extent of the domains of instability. In spite of that, physical
intuition and insight have led \cite{bh} to conclude that $\beta\approx 1$ is the lower limit for
the disk thickness for the occurrence of the classical MRI in disks as for smaller values of $\beta$
the thickness of the disk is smaller than the wave length of the dominant unstable mode.
In the current work the stabilizing effects of the boundaries are taken explicitly into account and hence
the threshold thickness of the disk may be calculated easily from the WKB solutions obtained in Sec.2.
Furthermore, the number of unstable modes as a function of $\beta $ may be estimated with the aid of that solution.
Thus, for example, it may be shown that there are only three unstable modes within the disk as long as the its beta value is less than $ \sim 15 $.
Such information is significant for the study of the consequent nonlinear development of the instability.
In addition, the shape of the unstable (as well as the stable) perturbations has been obtained explicitly
and as may be seen by its expression (\ref{wkbsol}) may differ significantly from the plain waves
assumed in the classical works on MRI's. It is finally interesting to note, that going to
the limit of thick disks (see Section 2.4) and assuming a localized perturbation results in a
reduced rate of growth in comparison to the less realistic monochromatic classical result [see Eq.(\ref{longtime})].
Recently, \cite{cc}, and \cite{ck} have considered the effects of the axial localization of the
perturbations by studying axisymmetric ballooning modes in finite disks. Such modes are characterized by
finite values of the radial wave vector $k_r$. The appropriate turning points within the disk were
found and consequently a discrete spectrum of eigenfunctions and eigenvalues was obtained.
The growth rates were found to be smaller than their "long-cylinder" counterparts. In particular,
it is shown there that the $k_r\rightarrow 0$ limit cannot be obtained from their scalings and asymptotics,
as no turning points exist within the disk in that limit. In that sense the present work is complementary to
\citep{ck} as it analyzes that very limit  $k_r\rightarrow 0$. Indeed, under the current
scaling and asymptotic expansion turning points are found within the disk and the corresponding
discrete spectrum is obtained.

The linear MHD equations have been employed in order to study the
stability and time behavior of a rotating stratified Keplerian disks whose
density decreases with height. The full solution of the dynamical evolution in time of an localized Gaussian
wave packet is explicitly derived and it's long
time behavior is discussed. It is proven analytically (WKB) that MRI can be suppressed
in sufficiently thin disk $\beta < \beta_*\approx \pi/11$ (\cite{ms} observed from
numerical simulations that if $\beta\approx 1$ the disk becomes MRI stable). Numerical solutions
of the Cauchy problem however, indicates that $\beta_*$ is of order 1. This result is consistent
with values obtained by \cite{sm}.
For thicker disks, the number of discrete unstable mode
increases with the thickness of the disk.
However, due to the localization in real space the initial amplitude of
the unstable mode is diminishingly small and the time it outgrows the original wave packet is
significantly longer than its predicted inverse growth rate. Thus, it has been demonstrated
numerically the growth time of a perturbation that is initially localized within the inner 5\%
of the disk, may be an order of magnitude longer than the inverse growth rate of the fastest
growing unstable mode.

The considering of the boundary effects may have a large impact on the design
of laboratory experiments to model MRI, where a magnetic field have to be
quite strong and devices cannot be very high.

\section*{Acknowledgments}

The authors are greatly indebted to Oded Regev and Orkan Umurhan for fruitful discussions and insights,
as well as for their encouragement.

\appendix

\section[]{Orthogonality}
Both sides of the homogenous equation $L[\widetilde{b}_{\varphi, n}(z)]=0$ are first multiplied by
$\widetilde{b}_{\varphi, m}(z)$ and then integrated over the interval $(-\infty,\infty)$.
Integrating by parts and setting the boundary terms to zero due the homogeneous BC (\ref{mhdbcinh})
yields
\begin{equation}
\int^\infty_{-\infty}\widetilde{b}_{\varphi, n}\left[\left(e^{z^2}\left(e^{z^2}
\widetilde{b}'_{\varphi,m}\right)''\right)'+3\beta\left(e^{z^2}\widetilde{b}'_{\varphi, m}\right)'\right]dz + 2\lambda^2_n
\int^\infty_{-\infty}\widetilde{b}_{\varphi, m}\left(e^{z^2}\widetilde{b}'_{\varphi, n}\right)'dz +
\lambda^2_n(\lambda^2_n-\beta)\int^\infty_{-\infty}\widetilde{b}_{\varphi, m}\widetilde{b}_{\varphi,n}dz=0
\label{a1}
\end{equation}
Defining
$$
Q_{1,m,n}\equiv\int^\infty_{-\infty}\widetilde{b}_{\varphi, m}\widetilde{b}_{\varphi, n}dz, \quad
Q_{2,m,n}\equiv\int^\infty_{-\infty}\widetilde{b}_{\varphi,m}\left(e^{z^2}\widetilde{b}'_{\varphi,n}\right)'dz=
\int^\infty_{-\infty}\widetilde{b}_{\varphi, n}\left(e^{z^2}\widetilde{b}'_{\varphi, m}\right)'dz,
$$
and noticing that to each eigenfunction $\widetilde{b}_{\varphi, n}$ correspond two eigenvalues $\lambda^2_{n,1,2}$
allows rewriting (\ref{a1}) in the form of the following set of homogenous linear equations:
\begin{equation}
A{\bf Q}=0, \quad {\bf Q}\equiv (Q_{1,m,n}, Q_{2,m,n})^T, \quad A\equiv\left (\begin{array}{cc}
\lambda^2_{n,1}(\lambda^2_{n,1}-\beta)-\lambda^2_{m}(\lambda^2_{m}-\beta) & 2(\lambda^2_{n,1}-\lambda^2_{m})
\\
\lambda^2_{n,2}(\lambda^2_{n,2}-\beta)-\lambda^2_{m}(\lambda^2_{m}-\beta)&2(\lambda^2_{n,2}-\lambda^2_{m})
\end{array} \right )
\end{equation}
Implying that different eigenfunctions ($n\neq m$) have different eigenvalues results in:
\begin{equation}
det(A)=2(\lambda^2_{n,1}-\lambda^2_{n,2})(\lambda^2_{n,1}-\lambda^2_{m})(\lambda^2_{n,2}-\lambda^2_{m})\neq 0.
\end{equation}%
This implies that
\begin{equation}
Q_{1,m,n}=Q_{2,m,n}=0.
\label{a4}
\end{equation}
In the opposite case of identical eigenfunctions ($m=n$) the equality $det(A)=0$ means that $Q_{1,n,n}\neq 0$.
Combining the latter with eq.(\ref{a4}) gives
\begin{equation}
\langle\widetilde{b}_{\varphi,m},\widetilde{b}_{\varphi,n}\rangle\equiv Q_{1,m,n}\equiv\int^\infty_{-\infty}\widetilde{b}_{\varphi, m}
\widetilde{b}_{\varphi,n}dz=\|\widetilde{b}_{\varphi,n}\|^2\delta_{m,n}.
\end{equation}%

\section[]{Green's function}
To obtain the expansion coefficients $c_n$ of a solution $\widetilde{b}_{\varphi}(z)$ for the inhomogeneous
equation $L[\widetilde{b}_{\varphi}(z)]=-f(z)$ in terms of the eigenfunctions $\widetilde{b}_{\varphi,n}(z)$
the solution expansion $\widetilde{b}_{\varphi}(z)=\sum_l c_l \widetilde{b}_{\varphi,l}(z)$ and expansion of
the r.h.s of the inhomogeneous equation $f(z)=\sum_m d_m \widetilde{b}_{\varphi,m}(z)$ is substituted into
the inhomogeneous equation. The result is multiplied by $\widetilde{b}_{\varphi, n}(z)$ and integrated over the
interval $(-\infty,\infty)$. The result is
\begin{equation}
\sum_lc_l\int^\infty_{-\infty}\widetilde{b}_{\varphi,n}L[\widetilde{b}_{\varphi,l}]=-d_n\|\widetilde{b}_{\varphi,n}\|^2.
\label{b1}
\end{equation}%
Defining in addition to $Q_{1,n,l}$ and $Q_{2,n,l}$
$$
Q_{3,n,l}=\int^\infty_{-\infty}\widetilde{b}_{\varphi,n}\left(e^{z^2}\left(e^{z^2}
\widetilde{b}'_{\varphi,l}\right)''\right)'dz,
$$
and employing the homogenous equation $L[\widetilde{b}_{\varphi,l}]=0$ multiplied by $\widetilde{b}_{\varphi, n}(z)$
and integrated over the interval $(-\infty,\infty)$ yields:
\begin{equation}
Q_{1,n,l}=\|\widetilde{b}_{\varphi,l}\|^2\delta_{n,l}, \quad
Q_{2,n,l}=-\frac{\|\widetilde{b}_{\varphi,l}\|^2}{2}(\lambda^2_{l,1}+\lambda^2_{l,2}-\beta)\delta_{n,l}, \quad
Q_{3,n,l}=\frac{\|\widetilde{b}_{\varphi,l}\|^2}{2}\left[3\beta(\lambda^2_{l,1}-
\beta)+\lambda^2_{l,2}(2\lambda^2_{l,1}+3\beta)\right]\delta_{n,l}.
\end{equation}%
Thus eq.(\ref{b1}) can be rewritten as
\begin{equation}
c_n(\lambda^2_{n,1}-\lambda^2)(\lambda^2_{n,2}-\lambda^2)=-d_n,
\end{equation}%
and a solution of the inhomogeneous equation with r.h.s $-f(z)$ has a form:
\begin{equation}
\widetilde{b}_{\varphi}(z)=-\sum_n\frac{d_n\widetilde{b}_{\varphi, n}(z)}{(\lambda^2_{n,1}-\lambda^2)(\lambda^2_{n,2}-\lambda^2)}.
\label{b4}
\end{equation}%
Finally noticing that $d_n=\frac{1}{\|\widetilde{b}_{\varphi,n}\|^2}\int^\infty_{-\infty}f(z)\widetilde{b}_{\varphi, n}(z)dz$,
a solution of inhomogeneous equation (\ref{b4}) can be rewritten as
\begin{equation}
\widetilde{b}_{\varphi}(z)=-\int^\infty_{-\infty}G(z,\zeta)f(\zeta)d\zeta,
\end{equation}%
where
$$
G(z,\zeta)=\sum_n\frac{\widetilde{b}_{\varphi,n}(z)\widetilde{b}_{\varphi,n}(\zeta)}{(\lambda^2_{n,1}-\lambda^2)
(\lambda^2_{n,2}-\lambda^2)\|\widetilde{b}_{\varphi,n}\|^2}
$$
is a Green's function written in terms of an eigenfunctions $\widetilde{b}_{\varphi,n}(z)$.

\bsp
\label{lastpage}
\end{document}